# Markets are efficient if and only if $P = NP$

*Philip Maymin, NYU-Polytechnic Institute*

*phil@maymin.com*

## ABSTRACT

I prove that if markets are weak-form efficient, meaning current prices fully reflect all information available in past prices, then $P = NP$, meaning every computational problem whose solution can be verified in polynomial time can also be solved in polynomial time. I also prove the converse by showing how we can "program" the market to solve $NP$-complete problems. Since $P$ probably does not equal $NP$, markets are probably not efficient. Specifically, markets become increasingly inefficient as the time series lengthens or becomes more frequent. An illustration by way of partitioning the excess returns to momentum strategies based on data availability confirms this prediction.

Are the concepts of market efficiency in the field of finance and computational efficiency in the field of computer science really the same thing?

The efficient market hypothesis claims that all information relevant to future prices is immediately reflected in the current prices of assets. In other words, you cannot consistently make money using publicly available information.

Specifically, the weakest form of the EMH states that future prices cannot be predicted by analyzing prices from the past. Therefore, technical analysis cannot work in the long run, though of course any strategy could make money randomly. Most finance academics believe markets are weak form efficient: Doran, Peterson, and Wright (2007) survey more than 4,500 finance professors and find that of the nearly 650 usable responses, the majority believe the US stock market is weak form efficient; only 8 percent generally disagree.

Computational efficiency in computer science distinguishes two kinds of algorithms. One class of algorithms belong to the set known as *P*, short for Polynomial, because they can *find* a solution to an input of length *n* in a time that is polynomial in *n*. For example, finding a sorted version of an input list can be done in polynomial time. Another class of algorithms belong to the set known as *NP*, short for Nondeterministic Polynomial, because they can *verify* a proposed solution to an input of length *n* in a time that is polynomial in *n*. Verifying a solution means that the algorithm always halts and answers whether the given proposal really does satisfy the input. For example, is it possible to visit each capital city in Europe exactly once and within a week? Given a proposed solution, verifying it can be done in polynomial, indeed linear, time. Does the proposed route visit each capital? Is the total travel time less than a week? (Finding a satisfying solution, however, might only be possible in exponential, not polynomial, time. This problem is known as the traveling salesman problem.) Polynomial time is, in short hand, considered efficient.

Obviously *P* is a subset of *NP*: any algorithm that can efficiently generate a solution can efficiently verify a proposed solution. The outstanding question in computer science is: does *P* = *NP*?

In other words, if a solution can be verified efficiently, does that mean it can be computed efficiently? Can the traveling salesman and similar problems be solved efficiently?

Consider the problem of satisfiability. Given a Boolean expression comprised of variables and negated variables combined through ANDs and ORs, where each variable can be either TRUE or FALSE, it is easy to confirm whether a particular proposed solution in fact satisfies the formula, but in general finding the solution seems to require checking every possible assignment of variables to TRUE or FALSE. For *n* variables, that requires searching through $2^n$ possibilities, meaning the algorithm works in exponential time, not polynomial time.

However, there has been no proof that there does not exist some algorithm that can determine satisfiability in polynomial time. If such an algorithm were found, then we would have that $P = NP$. If a proof is discovered that no such algorithm exists, then we would have that $P \neq NP$.

Just as most people in the field of finance believe markets are at least weak-form efficient, most computer scientists believe $P \neq NP$. Gasarch (2002) reports that of 100 respondents to his poll of various theorists who are "people whose opinions can be taken seriously," the majority thought the ultimate resolution to the question would be that $P \neq NP$; only 9 percent thought the ultimate resolution would be that $P = NP$.

The majority of financial academics believe in weak form efficiency and the majority of computer scientists believe that $P \neq NP$. The result of this paper is that they cannot both be right: either $P = NP$ and the markets are weakly efficient, or $P \neq NP$ and the markets are not weakly efficient.

## Literature Review

Efficiency was debated even before it was formulated in its present state. Simon (1955) argued that investors are only bounded rational. This paper shows that that even if that bound grows polynomially, due to advances in technology or human capital, it would not be enough to compensate for the exponential growth in the possible strategies as the length of the data increases.

Several researchers have argued that the strong form of the efficient market hypothesis is impossible. For example, Grossman (1976), Grossman and Stiglitz (1980), and Lo and MacKinlay (1999) argued that prices cannot perfectly reflect all available information because information is costly, and perfectly efficient markets would mean the rewards for gathering information were zero, so no trader would undergo the task. Campbell, Lo, and MacKinlay (1997) further note that both information-gathering and information-processing are expenses investors pay to achieve higher profits. This paper focuses instead on the weak form of the efficiency argument and the information-processing costs as applied solely to analyzing past prices, links it to the concept of computational efficiency, and makes specific predictions about the relation between the inefficiency and the amount of data available.

Daskalakis, Goldberg, and Papadimitriou (2009) and Monroe (2009) discuss possible complexity classifications of finding the equilibria of games or entire economies, where the question of whether $P = NP$ ultimately determines whether or not markets clear; in short, whether or not the participants are able to compute the resulting market prices from their preferences in standard utility models. Here, we assume markets do clear, and market prices do exist, whether or not standard utility models hold, and seek only to determine whether the real market can be efficient or not, conditional on the question of $P = NP$.

Arora, Barak, Brunnermeier, and Ge (2009) employ the tools of computational complexity to show that pricing a particular kind of derivative can be so computationally inefficient as to amplify the costs of asymmetric information. Instead of considering the difficulty of a valuation model, this paper considers the difficulty of searching trading strategies on an arbitrary financial instrument, whether it is a security or a derivative.

Chen, Fortnow, Nikolova, and Pennock (2007) show that finding arbitrage opportunities among wagers in a pair betting system of the type “horse A will beat horse B” is *NP*-hard. In other words, evaluating a set of orders from a collection of traders for the existence of a multi-party riskless match is computationally difficult. Here, we explore a different question: can a single investor, not multiple participants, explore all possible strategies based on historical prices, not orders, to find one that works? Their result is most similar to the second part of this paper in which we assume an efficient market and attempt to “program” it to solve an arbitrary *NP*-complete problem, because the way in which we attempt to encode an *NP*-complete problem into the market involves market orders.

Hasanhodzic, Lo, and Viola (2009) propose a framework for studying the efficiency of markets through a sequence of increasingly larger resources, such as time or memory. Specifically they consider strategies that depend only on a few of the most recent observations, similar to the way we will explore them here. But while they explore the ramifications onto the asset prices of such a world, and show that it can lead to bubbles, and that longer memory can result in greater arbitrage opportunities, this paper explores the exponentially increasing difficulty of finding optimal trading strategies with longer memories, and provides the identification between market efficiency and computational efficiency.

The key element in definitions and classifications of market efficiency revolve around information. The traditional classification of the different types of market efficiency into weak form (past prices), semi-strong form (all publicly available information), and strong form (private information) was updated by Fama (1991) to instead focus on tests of return predictability, event studies, and tests for private information. It is easiest to understand the results of this paper as applying to the traditional weak form of market efficiency, a subset of the newer classification of tests of return predictability that allows not only the use of past prices, but other past variables such as dividend yields, interest rates, and market capitalization.

Starting from Shannon (1948) and Shannon and Weaver (1949), information was understood through the concept of the Shannon entropy, the uncertainty remaining in the message and the limits of lossless compression. Weinberger (2002) reviews the information research to date and notes that little attention has been paid to the effectiveness of the information to influence the receiver's actions. To fill this gap, Weinberger (2002) introduces the theory of pragmatic information as the change in the probabilities of the receiver's actions, thus making the effect of information subjective and relative to the receiver; in those terms, a market is efficient if the pragmatic information of previous price histories is zero. These alternative viewpoints of information do not affect the results of this paper, as we are essentially showing the difficulty of computing all of the implications from the static history of past prices.

The thrust of this paper is that the information cost of searching all possible patterns in a sequence of prices is an exponential task, so as the amount of data gets large, at some point it will overwhelm the aggregate ability of all investors to discover patterns, and therefore there should be positive returns to those who do find the patterns, at least until those patterns become widely known. This is exactly the phenomenon widely discovered in the literature. Tóth and Kertész (2006) found evidence of increasing efficiency in the New York Stock Exchange as measured by the disappearance of a lead-lag relationship after that relationship was publicly documented by Lo and MacKinlay (1990), and G. William Schwert (2003) found that anomalies in general tend to substantially weaken after publication of the research findings and the subsequent trading by practitioners. These findings are in line with what we would expect to see if evaluating all possible price patterns were computationally inefficient.

## Approximations and Alternatives

There are approximations to both market efficiency and computational efficiency. Computationally, many problems that are technically in *NP* can be solved quite quickly with approximations or randomization algorithms. Rodriguez, Villagra, and Baran (2008) show that modern solvers for satisfiability problems can combine different algorithms to interact with each other in a search for a global solution. Even if the worst-case performance is still exponential, many common cases can often be solved much faster.

Financially, the "economic calculation problem" of von Mises (1920) and Hayek (1935) suggests, among other things, that, even if a free market is not perfectly efficient, it will certainly be more efficient than a regulatory or government alternative. In other words, even if mispricings occasionally occur, most of the time they are smaller than any other alternative system.

Both of those arguments are similar in their domains but neither applies to the results of this paper. Whether markets are efficient or not, and whether $P = NP$ or not, there is no doubt that there will be markets that can allocate resources very close to efficiently and there will be algorithms that can solve problems very close to efficiently. The results of this paper should not be interpreted as support for government intervention into the market; on the contrary, the fact that market efficiency and computational efficiency are linked suggests that government should no more intervene in the market or regulate market participants than it should intervene in computations or regulate computer algorithms.

Furthermore, while computational performance often depends on problems and inputs that are extrinsic, and where average-case performance could be quite important, in financial markets, even knowing that 99 percent of all strategies do not work is not enough to conclude that markets are efficient. On the contrary, finding those remaining profitable strategies would be a paramount goal for many practitioners.

In short, approximations allow for practical computation to be quite fast even if it is not the case that *P* = *NP*, but the impact of those approximations are low in the financial space because of the enormous value in finding a profitable strategy.

An alternative scenario is one where *P* = *NP* but the order of the polynomial is so high as to be effectively as incomputable as an exponential. In this scenario, the theoretical impact is large as one of the deepest problems in computer science would be solved, but the practical impact is small, as it would not mean any substantial speedup in any valuable algorithms. Would such a scenario preclude the results of this paper?

Not at all. If *P* = *NP*, even with a high exponent on the polynomial, that means that checking strategies from the past becomes only polynomially harder as time passes and data is aggregated. But so long as the combined computational power of financial market participants continues to grow exponentially, either through population growth or technological advances, then there will always come a time when *all* past strategies can be quickly backtested by the then-prevailing amount of computational power. In short, so long as *P* = *NP*, the markets will ultimately be efficient, regardless of how high the exponent on the polynomial is; the only question is when, and the answer depends on the available computational power. For example, even a hundred years ago the combined computational power of all investors may not have been enough to find even a simple anomaly, even if all *NP* problems were linear, i.e. polynomial of order one.

This paper is structured as follows. First, we explore the financial implications if *P* = *NP* to find that at least one implication of such a monumental assumption is that markets would indeed be efficient. Next, we show that efficient markets, when combined with some more minor, technical assumptions, implies that *P* = *NP*. Therefore, markets are efficient if and only if *P* = *NP*. What are the currently testable implications of this theory? We test some predictions in Section III. Finally, we conclude, calling for a more rigorous, algorithmic definition of efficient markets going forward.

## I. The Information in Past Prices

What is *information* in the context of a financial time series? Suppose the time series of past prices, sufficiently analyzed, suggests a particular pattern, for example that three UP days tend to be followed by another UP day far too often. The EMH asserts two claims: 1) as soon as this pattern is found, the effect should vanish, as people start trading earlier and earlier to attempt to profit from the pattern, and 2) this pattern will be found immediately. Barberis and Thaler (2003), among others, refer to these respectively as the "no free lunch" and the "price is right" assertions of the EMH.

This paper takes no stand on the "no free lunch" assertion, other than to note that it requires that the particular predictable pattern must be public knowledge, e.g. through the publication of an academic paper or common industry practice. If only a few investors know of and trade the effect with little market impact, they could in fact profit consistently.

This paper does suggest, however, that the “price is right” assertion of the EMH is true only if all patterns can be efficiently (i.e. quickly) evaluated, which can happen either if there is not too much data, or if $P = NP$ so that any efficiently verifiable predictable pattern is therefore efficiently computable as well.

The basic argument is as follows. For simplicity but without loss of generality, assume there are *n* past price changes, each of which is either UP (1) or DOWN (0). How many possible trading strategies are there? Suppose we allow a strategy to either be long, short, or neutral to the market. Then the list of all possible strategies includes the strategy which would have been always neutral, the ones that would have been always neutral but for the last day when it would have been either long or short, and so on. In other words, there are three possibilities for each of the *n* past price changes; there are $3^n$ possible strategies.

We ask our question, call it question 1: “Does there exist a strategy that statistically significantly (after accounting for possible data mining) makes money (after accounting for transactions costs)?” An answer to that question essentially tests each of the possible $3^n$ strategies.

Given a particular strategy, of course, it would merely require one linear pass through the data to determine whether the profitability exceeds random chance, so the verification is $O(n)$, i.e. it takes some constant amount of time multiplied by the length of the time series *n*; if *n* doubles, it will take twice as long. Therefore the decision problem of finding a profitable strategy is in *NP*, because verifying a candidate solution can be done in polynomial time.

But given that there are $3^n$ patterns to test, finding a solution, as opposed to verifying one, requires $O(3^n)$, i.e. it is exponential in the size of *n*. For small *n*, this is computable, even though it is exponential. But as *n* grows, it becomes impossible to check every possible pattern quickly.

But what if there are *M* people, each of whom is checking some random subset of the possible patterns? Then the inevitable is merely delayed; once the time series gets long enough, there will always come a time when the number of possible patterns exceeds even the computational ability of all humans and computers on earth, assuming that $P \neq NP$.

Therefore, as asset histories grow, whether through the passage of time or the introduction of finer data (daily, intraday, tick data), the market should become less efficient. Longer time series of data will provide more anomalies.

On the other hand, if $P = NP$, then the efficient market hypothesis will hold, and future prices will be unpredictable from any combination of past prices, because then even a single investor will be able to check all possible paths relatively quickly. Instead of exponential time, there will exist some algorithm capable of checking all of the possible patterns in polynomial time. In that world, virtually all computation would be efficient, and so would the market.

The above discussion gives a flavor of the argument, and much of the intuition, but one drawback of examining all $3^n$ possible strategies is that there will always exist one that generates a total profit of *n*, namely, the one that is long on days the market went UP and short on days the market went DOWN. The problem is that we have not formally defined a strategy, and the definition we used above was too lenient. Not every collection of buy and sell signals is a bona fide strategy; a true strategy must generate buy and sell signals deterministically given a past pattern. That is what it means to have a consistent strategy going forward.

## Definition: Technical Strategy

Let us say that a *technical strategy* is a function *S* that takes a sequence of *t* UPs and DOWNs encoded as 1's and 0's and outputs a position, either +1 for long, -1 for short, or 0 for neutral:

$$S: (r_i \in \{0,1\})_{i=1}^{t} \rightarrow \{-1,0,+1\}$$

Let us refer to those strategies *S* that never output -1 as *long-or-out technical strategies*. They are always either long the market or not in the market. This is merely for convenience; such strategies are easier to think about but the assumption does not affect the results.

Note that for each particular technical strategy *S*, the length *t* of the sequence it takes as input is fixed. Thus, it is not the case that as we progress in time, the strategy is given the entire history of past prices, an ever-increasing input. On the contrary, the strategy is always only given the most recent *t* prices (or in our case, the most recent *t* directions of price changes, UP or DOWN). Furthermore, the input is always on a sliding basis: the first time the strategy is called with inputs from $i = 1$ through $i = t$; the next time with inputs from $i = 2$ through $i = t + 1$; and so on. If, instead, $t$ were allowed to increase, then we would still allow any arbitrary strategy, including the one that happened to always buy on the UP days and sell on the DOWN days, because the outcome could be coded to depend on the length of the series being input. (Alternatively, if the strategy is computationally constrained in memory usage or distinct states, this assumption can be relaxed so long as the strategy still only depends on past prices, and the results will still hold.)

The idea of having a fixed lookback window also accords both with our intuitions of what a strategy ought to be and what it is in practice. Intuitively, a strategy ought to be a kind of black box that will do the same thing if the same circumstances hold in the future. In practice, the most popular technical strategies are variants of momentum strategies, which have fixed lookback windows of some number of months, following Jegadeesh and Titman (1993) and the literature that followed.

Now our question 1 transforms into the more formal question 2: "For a given lookback window $t$, does there exist a long-or-out technical strategy $S$ that generates a statistically significant profit on the historical time series of length $n$?"

There is no way to answer this question given only the data. Statistical significance implies probabilities, and probabilities imply hypothesis testing. Just because some strategy produced some profits in the past, does not by itself mean that we can repeat it successfully. We need to have a model of the time series; indeed, we technically need to have two models, a null hypothesis and an alternative hypothesis, before we can determine whether a particular given strategy truly generates statistically significant profit.

This is just another formulation of the well-known joint hypothesis problem of Fama (1970). As e.g. Campbell, Lo, and MacKinlay (1997) put it, "any test of efficiency must assume an equilibrium model that defines normal security returns."

We can rewrite our question in terms of a given model of market equilibrium, and an assumed alternative model for the purposes of hypothesis testing, by observing that the effect of such choices will serve only to define the critical value $K$ that separates statistically significantly profitable strategies from the rest. This choice of the critical value essentially incorporates into it the particular equilibrium model assumptions.

We now ask question 3: "For a given lookback window $t$ and a critical value of profit $K$ (determined from given null and alternative models of market equilibrium) does there exist a long-or-out technical strategy $S$ that generates a profit in excess of $K$ on the historical time series of length $n$?"

In other words, we can allow arbitrary models of market equilibrium because all we care about is the implied critical profit $K$.

One debate among both academics and practitioners is how to reasonably search among possible strategies. Some might argue that only strategies based on economic justification, for example behavioral biases, are reasonable to explore. Others might use blind data mining, for example through machine learning or neural networks, to find strategies without any clear economic justification. Must we take a stand on the statistical nature of truth in order for the results of this paper to hold?

It turns out that we do not. All such assumptions form an implicit part of a model of market equilibrium and therefore merge into the choice of the critical value.

For some models, $K$ will equal $n$, meaning even the best possible strategy, the one that generates the maximal possible profit on the given time series, will still not be statistically significantly profitable relative to the given models.

On the other hand, there will also be models that do generate a value of $K$ that is substantially less than $n$, so that it is indeed possible for some always-in technical strategy $S$ with a lookback window of $t$ to generate enough profit.

In those instances, how can we find such a strategy $S$? Alternatively, how can we be sure no such $S$ exists? Verifying the profit of a given strategy on the time series is easy and can be done in linear time, but there are $2^{t+1}$ possible strategies (because every possible strategy is a mapping from $2^t$ different possible sequences of 1's and 0's of length *t* into two possible choices, LONG or OUT), so testing them all would seemingly require time exponential in *t*.

In fact, we can answer question 3 in polynomial time with the following algorithm. Consider any *t*-length subseries of the data, e.g. (0, 1, 1, 0, …, 1). We can interpret this as an integer written in base two; suppose it is 137. Then we gather all of the subsequent returns that occur for each unique *t*-length subseries; for example the result might look like {(137, {1, 0, 1}), …}, meaning that that particular *t*-length subseries occurred three times, and was followed by positive returns twice and a negative return once. Take the sum of all of the returns; in this case the sum of the returns associated with 137 is 2. Then the following strategy is the best possible one: for each subseries, if the associated total return is positive, map the subseries to being long (+1), and if it is negative, map the subseries to being out (0). This will generate the maximum possible profit, which we then compare with *K* to see if it exceeds our target.

So question 3 turned out to be too simple. If that was the only question investors were interested in asking, and there was only one asset generating a time series, then the market could be efficient even if $P \neq NP$, because there is a polynomial time algorithm for determining the answer.

But in fact there are multiple assets or securities, and the nature of a strategy is that it apply equally to any of them. In other words, a strategy *S* applied to a *t*-length subseries of asset 1 ought to output the same position recommendation when the same strategy S is applied to a *t*-length subseries of asset 2.

Furthermore, having multiple assets potentially simultaneously requiring investment introduces a budget constraint. Let us assume that investors do not have access to leverage and must put up the full capital to purchase any asset they wish to buy.

Now we ask question 4: "For a given lookback window *t* and a critical value of profit *K* (determined from given null and alternative models of market equilibrium) does there exist a long-or-out technical strategy *S* that generates a profit in excess of *K* on a collection of *t*-length vectors of zeros and ones such that the sum of the prices of the purchased assets does not exceed a budget constraint *B*?"

Compare this question to the following question, known as the knapsack problem because it asks how to optimally choose items of different sizes and values to place into a bag:

**Knapsack question** (following Garey and Johnson (1979)): "Given a finite set *U*, and, for each $u \in U$, the two positive integers *s(u)* and *v(u)* which are referred to as *u*'s "size" and "value" respectively, and given two positive integers *B* and *K*, is there a subset $U' \subseteq U$ such that $\sum_{u \in U'} s(u) \leq B$ and such that $\sum_{u \in U'} v(u) \geq K$?"

The knapsack problem is *NP*-complete. It is also a rephrasing of our question 4, where we interpret the size *s(u)* as the price of the given security following a particular *t*-length subseries, the value *v(u)* as the future return of the asset following the particular subseries, and the subset $U'$ as the collection of *t*-length subseries which would result in a long purchase by the long-or-out strategy *S* (so that the complement $U - U'$ represents no position by *S*). In the analogy of the knapsack, we are trying to pick those *t*-length series for various assets where the combined profit is greatest, subject to our simultaneous budget constraint.

In other words, the question of whether or not there exists a budget-conscious long-or-out strategy that generates statistically significant profit relative to a given model of market equilibrium is the same as the knapsack problem, which itself is *NP*-complete. Therefore, investors would be able to quickly compute the answer to this question if and only if *P* = *NP*.

Note, however, that as the time series of price changes doubles in length, if the lookback window of the strategies we are exploring remains constant, the set *U* will not change once it includes all possible *t*-length subseries. In light of this, the final assumption that we add is that *t* is proportional to *n*; in other words, as the length of the time series of price changes doubles, investors care about testing strategies with double the lookback window. This assumption is also sensible because investors may view a given time series of length *n* as e.g. 20 non-overlapping sequences of length *n*/20, and wish to test all strategies of length *t* = *n*/20. Hence, as *n* doubles, so too does *t*. The only technical requirement is that *t* grows as *n* grows, so that *U* grows either as *n* grows, or as we add more assets. Therefore, if *P* = *NP* and we are able to test all possible strategies computationally efficiently, we can then trade based on our results to make the market efficient as well.

In short, if $P \neq NP$, then the market cannot be efficient for very long because the ability of investors to check all strategies will be quickly overwhelmed by the exponential number of possible strategies there are to check. Contrariwise, if indeed $P = NP$, then the market will be efficient because all possible strategies can be quickly checked in polynomial time.

Therefore, the market is efficient if and only if $P = NP$.

## II. Programming the Market

It is clear from the previous section that $P = NP$ implies market efficiency, but perhaps it is a bit unsettling that market efficiency implies $P = NP$; after all, that would mean that any arbitrary difficult problem in *NP* could be solved by particular machinations of the market, and we have not provided the algorithm for those machinations explicitly. We do so in this section.

Assume that the weak form of the EMH holds for all *n*. Can we use market efficiency to obtain computational efficiency? For the purposes of this section, we will explicitly suppose that the past information that is fully reflected in current prices includes not only past prices, but also all past publicly available order histories. Essentially we assume that the order book for each security is as easily and publicly available as the actual trades. By itself, this is a reasonable assumption in modern markets. For example, the NYSE's TAQ database lists all intraday quotes, including best bid and offer prices and sizes.

But we need to make one other assumption that is currently not standard in modern markets: we need to allow participants to place order-cancels-order ("OCO") or one-or-the-other orders. These are orders on different securities that automatically cancel the remaining orders whenever one is hit. An example is an order to buy one share of ABC at 100 or sell two shares of XYZ at 40, order-cancels-order. Then, as soon as the market is such that one of those two orders is filled, the other is cancelled automatically. There can be no chance that both orders are filled. This functionality is not currently available across different securities (it is available within a single security, e.g. buy at a particular price or sell at another price, order-cancels-order). Furthermore, we must allow such OCO orders to include not just two, but up to three securities. Call this assumption OCO-3.

This OCO-3 assumption is not currently available but there is no reason to believe it would continue to be impossible or impractical several decades from now. Thus, in principle, the implications of this section on programming the market will be testable when the market evolves and allows more sophisticated order handling abilities.

The problem we will attempt to encode into the market is that of satisfiability. As shown by Cook (1971) and Levin (1973), this problem is *NP*-complete, meaning that any other *NP* problem can be expressed as an instance of satisfiability. In other words, a polynomial solution to satisfiability provides a polynomial solution to any *NP* problem, and so proves *P* = *NP*. The thrust of this section is to show how an assumedly efficient market (which also allows OCO-3 orders) can solve satisfiability quickly.

Satisfiability is a decision problem that takes an input composed of three elements: variables and negated (!) variables, conjunctions (AND), and disjunctions (OR), all possibly grouped by parentheses. An example input might be (*a OR b) AND !c*. The input is deemed satisfiable if there exists a mapping of each occurring variable into TRUE or FALSE such that the overall expression is TRUE. In this example, $\{a \to \mathrm{TRUE}, c \to \mathrm{FALSE}\}$ and $\{b \to \mathrm{TRUE}, c \to \mathrm{FALSE}\}$ each satisfy the expression, so this input would be satisfiable. An input such as *a AND !a* is not satisfiable.

A particular form of satisfiability is called 3-SAT and it operates only on inputs that are in "conjunctive normal form," meaning parentheses group collections of three literals (a literal is a variable or a negated variable) combined with OR's, and those collections are themselves combined with AND's. An example might be:

*(a OR b OR !c) AND (a OR !b OR d)*

Karp (1972) shows that 3-SAT is itself NP-complete, and it remains NP-complete even if the number of solutions is guaranteed to be either zero or one (Valiant and Vazirani (1985)), so we will focus our attention on trying to use the efficiency of the market to decide a 3-SAT problem.

Here is how we will program the market to solve an arbitrary 3-SAT problem. We will interpret each variable as a security. Bare variables imply buy orders, and negated variables imply sell orders. Each collection of three variables will be an OCO-3 order. Thus, for the 3-SAT example above, we will interpret it as two orders, to be placed simultaneously:

1. Order-cancels-order buy A, buy B, or sell C.
2. Order-cancels-order buy A, sell B, or buy D.

Note that the two OCO orders are separate: for instance, if we buy D, then the buy A and sell B orders of the second OCO order are cancelled, but the first OCO remains open until we either buy A, buy B, or sell C.

The size of each order will be the minimum lot size for each security, typically one hundred shares in the current market for most shares on the NYSE.

The price of each order will be the mid, the midpoint between the prevailing bid and offer. Note that efficiency in the market does not imply infinite liquidity or rule out bid-offer spreads or transactions costs: market efficiency merely rules out the possibility of trading at a price that does not fully reflect all available past information.

Thus, absent a market move in any of the securities, our compound order will not generate a trade instantaneously; we are tightening the market in each security by either offering to sell below the prevailing ask or offering to buy above the current bid.

The strategy we will follow for a given input is to merely place the encoding orders in the market, wait some constant amount of time, and then cancel all outstanding orders and liquidate any positions we have. Because we are trading minimum lot sizes, the liquidation cost is minimal.

So what should the market do? If it is truly efficient, and there exists some way to execute all of those separate OCO orders in such a way that an overall profit is guaranteed, including taking into account the larger transactions costs from executing more orders, then the market, by its assumed efficiency, ought to be able to find a way to do so.

In other words, the market allows us to compute in polynomial time the solution to an arbitrary 3-SAT problem.

## III.Data and Discussion

To be clear, we are also not trying to determine if $P = NP$ or, for that matter, whether markets are efficient. The point of this paper is to show the link between these two previously disparate fundamental questions from different fields: we have shown that markets are efficient if and only if $P = NP$. Furthermore, the most likely scenario, given this link and the fact that the overwhelming consensus among academics appears to be that $P \neq NP$, is that markets are not efficient, even when expressed in its weakest form.

Now, one may argue that a weaker form of market efficiency as per Fama (1991) and Jensen (1978) is that patterns are exploited until the marginal revenue of further discovery equals the marginal cost of further search. A counterargument to that would be graduate students and other hobbyists and day traders searching for an edge: to them, the marginal cost is zero and at times negative because the value of the search itself, regardless of the outcome, is a positive learning experience for them, and the enormous potential payout, even if only reputational or signaling payout, relative to their alternatives is enough to make them want to search further. Therefore, even weakly efficient markets ought to fully reflect all available information in past prices.

If we concede that $P \neq NP$, and therefore the market is not efficient, we may be able to observe greater inefficiency in the market when there is greater data, given the exponentially increasing complexity of checking all possible strategies.

Could the value/growth, size, and momentum anomalies, among others, be expressions of this computational phenomenon? Specifically is it the case that there are more anomalies, or the existing anomalies are more profitable, when the time series is longer?

To check, we can explore the time-series variability of profitability in the standard momentum strategy of Jegadeesh and Titman (1993). Why the momentum anomaly instead of, say, the value anomaly? The prediction from the link between computational efficiency and market efficiency is that unknown anomalies become even harder to find with more data, like piling more hay onto a haystack containing one needle. But the key is that the anomaly must have been unknown. After all, simply retesting a known strategy on longer data is clearly fast.

Was the value strategy an unknown strategy? Probably not. After all, many investors throughout history have often tried to buy stocks that seemed cheap relative to their holdings, and such activity is at least a noisy proxy for the value strategy.

But the momentum strategy likely was unknown, because of its peculiar and very particular construction, so it is a better choice.

Each month, the Jegadeesh and Titman (1993) strategy essentially sorts all stocks in the CRSP database based on their performance over the past few months (typically six) and then purchases the top decile (the winners) and sells the bottom decile performers (the losers), holding the zero-cost portfolio for some number of months (also typically six).

As of the end of the month of December in 1972, the CRSP database began including Nasdaq stocks, more than doubling the number of stocks in the database. About four-fifths of the total number of (stock, date) pairs of data occurred from December 1972 through December 2008, even though that range accounted for only about two-fifths of the total date range. Thus, the prediction of the computational link to market efficiency is that the momentum strategy should be more profitable in the years following December 1972 than before.

Jegadeesh and Titman (1993) used a sample period from January 1965 to December 1989. We look at the original paper rather than their followup (Jegadeesh and Titman (2001)) or any of the other later literature they inspired to examine a period of time when the momentum strategy was presumed unknown; later periods are contaminated by the market's knowledge of the strategy's performance and results.

In their original paper, they report the results of their strategy for three periods, 1927-1940, 1940-1965, and 1965-1989. Table 1 lists their reported average performance of the six-month into six-month momentum strategies. It also lists the number of data points in the CRSP database for each of the breakpoint years; these numbers are calculated by counting the number of (stock, month-end date) data points in the database where the month-end data is less than or equal to December of the given year.

**Table I: Momentum Strategies.** The table below lists the cumulative return performance of the original Jegadeesh and Titman (1993) six-month formation period, six-month holding period momentum strategy returns across different periods (table references are to their paper), along with the number of unique (CRSP permno, month-end date) pairs in the CRSP monthly stock database through December of each given year, rounded to the nearest thousand.

| **PERIOD** | **PERFORMANCE** | **CRSP DATA** |
|---|---|---|
| 1927-1940 | -6.56%<br>(Table VIII, Panel A) | 1940: 125,000 |
| 1941-1965 | 3.65%<br>(Table VIII, Panel B) | 1965: 459,000 |
| 1965-1989 | 5.10%<br>(Table VII) | 1989: 1,753,000 |

As expected, the earliest periods with little data showed no abnormal returns, the later period with nearly quadruple the cumulative data showed positive returns, and the latest period with a further quadrupling showed even more positive returns.

Of course, computing and data processing technologies flourished during the last period under examination, so the number of strategies that could be checked by investors increased as well. Nevertheless, these results do provide an illustration of the prediction that *ceteris paribus* more data should lead to more anomalies.

## IV. Conclusion

Perhaps the most famous question in the field of finance is: "Is the market efficient?"

Perhaps the most famous question in the field of computer science is: "Does $P = NP$?"

The result of this paper is that these two questions are linked, and furthermore, the answers to the two questions must be the same: markets are efficient if and only if $P = NP$.

Market efficiency means that prices fully reflect all available past information. $P = NP$ is shorthand for the claim that any computation problem whose solution can be verified quickly (namely, in polynomial time) can also be solved quickly. For prices to fully reflect all available past information, investors must be able to compute the best strategies for them to follow given the data available. This problem turns out to be so difficult, that if it can be done quickly, then any difficult problem can be done quickly, and so the set *P* of all problems whose solutions can be computed quickly equals the set *NP* of all problems whose solutions can be verified quickly.

The prevailing viewpoint in the field of finance is that the markets are probably at least weakly efficient. The prevailing viewpoint in the field of computer science is that *P* probably does not equal *NP*. With the results of this paper, it is clear that both cannot be correct.

Given that *P* = *NP* is a mathematical problem that has been tackled repeatedly over many decades, and market efficiency is a claim about empirical data that requires a model of market equilibrium to test, it is probably the case that markets are not efficient.

However, the conclusion of this paper is not merely to take sides on either of the two debates, but to link the two literatures to refine our understanding of efficiency. Future research on market efficiency might be better served if it incorporated some concepts of computational efficiency. Specifically, future tests of market efficiency should include not only a model of market equilibrium, but a model of market computation.